\documentclass[aps,prl,reprint,preprintnumbers,superscriptaddress]{revtex4-1}
\usepackage{graphicx,longtable,mathrsfs,color,array}
\usepackage[hidelinks]{hyperref}
\usepackage[usenames,dvipsnames]{xcolor} 
\usepackage{amssymb,amsmath,mathtools,mathrsfs,slashed} 
\usepackage{epsfig,subfigure,placeins,float} 
\usepackage{booktabs,longtable,ctable,multirow} 
\usepackage{exscale,relsize} 
\usepackage[normalem]{ulem} 
\usepackage{enumerate}
\usepackage{times, mathptmx} 
\usepackage{color}
\usepackage{hyperref}
\usepackage{graphicx}
\usepackage{color}
\usepackage{graphicx,graphics}
\usepackage[utf8]{inputenc}

\newcommand{\be}{\begin{equation}}
\newcommand{\ee}{\end{equation}}
\newcommand{\bea}{\begin{eqnarray}}
\newcommand{\eea}{\end{eqnarray}}

\usepackage[stable]{footmisc}
\usepackage{color}

\def\dkmu2{\delta K_{\mu \nu}\delta K^{\mu \nu}}
\def\pmu2{  \phi_{\mu \nu}\phi^{\mu \nu}}

\providecommand{\abs}[1]{\lvert#1\rvert}

\begin{document}

\preprint{SISSA 02/2018/FISI}

\title{Emergent Dark Energy from Dark Matter}

\author{Takeshi Kobayashi}
\email{takeshi.kobayashi@sissa.it}
\affiliation{SISSA and INFN Sezione di Trieste, Via Bonomea 265, 34136 Trieste, Italy}
\author{Pedro G. Ferreira}
\affiliation{Astrophysics, University of Oxford, DWB, Keble Road, Oxford OX1 3RH, UK}

\begin{abstract}
We consider the cosmological dynamics  of a scalar field in a 
 potential with multiple troughs and peaks. We show that the dynamics
 of the scalar field will evolve from light dark matter-like behaviour
 (such as that of a light axion) to a combination of heavy dark
 matter-like and dark energy-like behaviour. We discuss the
 phenomenology of such a model, explaining how it can give rise to the
 cosmological constant, as well as how it can decouple the 
 dark sector densities between the time of recombination and today,
 for both the homogeneous background and perturbations. 
 The final form of the dark matter is axion-like, but
 with abundance and primordial isocurvature modes taking very different
 values from traditional, axionic, dark matter.
\end{abstract}


\maketitle

\noindent
\textit{Introduction:}
Scalar fields have played an important role in modern theoretical
cosmology. They have been at the heart of inflationary cosmology,
driving the accelerated expansion at early times
\cite{Martin:2013tda}. They have been invoked as a plausible candidate
for dark matter, in the form of an axion or axion-like particles
\cite{Arvanitaki:2009fg,Ringwald:2012hr,Marsh:2015xka,Hui:2016ltb}. And
they are the leading candidate for dark energy, replacing the cosmological constant as the dominant energy source at late times \cite{Copeland:2006wr,Clifton:2011jh}.

The standard approach has been to consider models which have some fundamental underpinning. Typically this involves choosing a potential $V(\phi)$ for the scalar field, $\phi$, with a simple analytic form, with one or a few minima, or with a certain degree of periodicity. The resulting dynamics is often relatively simple: the dynamics of the scalar field is either monotonic (used for inflation and dark energy) or oscillatory (used for dark matter). It would make sense, however, to countenance the possibility that $V(\phi)$ is rich, structured, with many different scales and minima. Such a complex potential can easily arise if one considers multiple scalar fields, or in higher-dimensional universe, with many extra dimensions and highly intricate topologies \cite{Sloan:2016kbc}. A particularly interesting analogy that can be considered is with spin-glasses where multiple minima can lead to rich dynamics and complex phenomena \cite{Soljacic:1999im}.
In this paper we will explore the possibility that, at late times, a cosmological scalar field is embedded in a theory which has a high degree of complexity and show that novel dynamics can emerge.

\begin{figure*}[t]
\centering  
  \includegraphics[width=\linewidth]{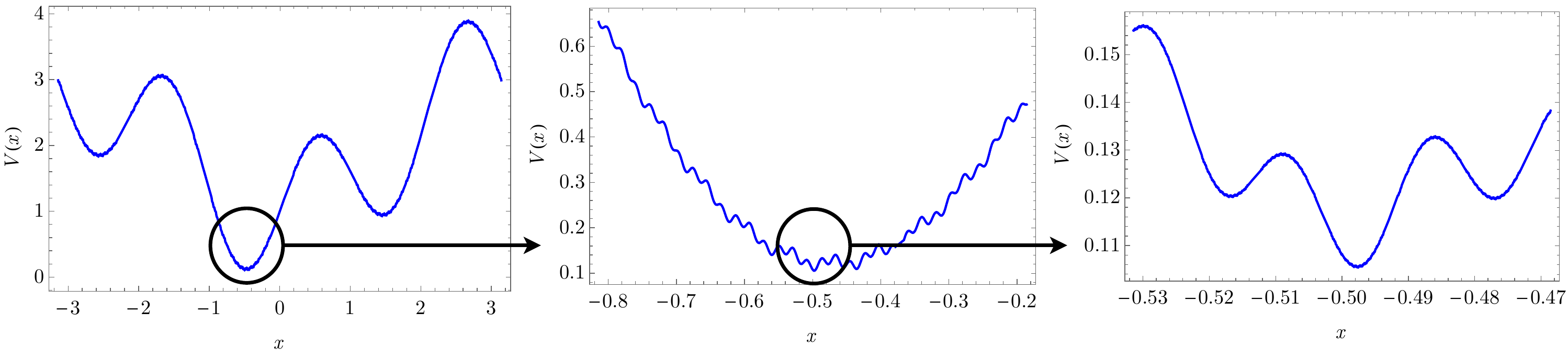}
  \caption{Cartoon of a scalar field potential with
  structure. As the universe expands and the scalar oscillation is damped,
  the scalar field `discovers' more local minima of the potential.}
  \label{fig:cartoon}
  \vskip -0.2in
\end{figure*}

We will consider a scenario with one degree of freedom,
i.e. one scalar field, that resides in an effective potential with
structures on different scales. 
Its origin may be in a multi-dimensional field space,
such as the string landscape or axiverse~\cite{Douglas:2006es,Arvanitaki:2009fg},
but for the purpose of this paper, we will model it as $V(\phi)$.
This can be viewed as focusing on the lightest direction in the
multi-field space.
It is also known that, under
certain symmetries, a multi-scalar field theory can relax to lower
dimensional dynamics (for an interesting example involving scale
symmetry, see \cite{Ferreira:2016vsc,Ferreira:2016wem}).
An example of the type of potential we are envisaging can be seen in
Figure \ref{fig:cartoon} where successive ``zoom ins'' of the potential
close to what looks like the global minimum reveals a rich structure of
local minima.
This ``zoom in'' naturally occurs for a cosmological scalar field oscillating along
the potential, since the expansion of the universe forces the oscillation to damp.
As a consequence, the oscillating scalar eventually gets trapped in one
of the local minima of the substructure.
We will show that this picture corresponds to a universe with dark
matter occasionally splitting into a mixture of dark matter plus dark energy,
and explore its phenomenological and cosmological consequences.

\noindent
\textit{Two-cosine model:}
Let us consider a minimally coupled real scalar field with an action
\begin{equation}
S=
\int d^4 x\sqrt{-g}\left[\frac{M^2_{\rm Pl}}{2}R
-\frac{1}{2}g^{\mu\nu}\partial_\mu\phi\partial_\nu\phi
-V(\phi)+L_{\mathrm{m}}\right] ,
\label{act}
\end{equation}
where $L_{\mathrm{m}}$ is the Lagrangian for other matter fields.
If we restrict ourselves to a homogeneous and isotropic space-time, with
$ds^2=-dt^2+a^2(t)(d\vec{r})^2$,
we arrive at the Klein-Gordon equation ${\ddot \phi}+3H{\dot \phi}=-V'$
and the Friedmann equation $ 3M^2_{\rm Pl} H^2 = \rho_\phi+\rho_{\mathrm{m}}$.
Here overdot is derivative with regards to~$t$, $H={\dot a}/a$,
$V'=dV/d\phi$, and $\rho_\phi =  V + \dot{\phi}^2 / 2 $ is the scalar
field energy density. 
Two salient regimes should be highlighted. If $V\simeq m^2\phi^2/2$ and
$m \gg H$, then $\phi$ will be oscillatory and $\rho_\phi\propto 1/a^3$;
the scalar field will evolve as a cold dark matter component.
If $V\simeq V_0$ and $V_0 \gg {\dot \phi}^2$, then $\rho_\phi$ will play
the role of a cosmological constant and, if further
$\rho_{\mathrm{m}}\ll\rho_\phi$ we have $H\simeq$ constant.

As the simplest potential that exhibits structures on different scales,
let us consider a potential consisting of two cosines:
\begin{equation}
V(\phi)=V_0+ m^2 f^2 \left\{1-\cos\left(\frac{\phi}{f}\right)
   +c\left[1-\cos\left(n\frac{\phi}{f}+\delta\right)\right]\right\}.
\label{2cos}
\end{equation}
Here $V_0$ is an offset with mass dimension four, $m$ and $f$ are mass
scales, and $c$, $n$, and $\delta$ are dimensionless. 
We choose $m, f, c > 0$ and $n > 1$.
The $\cos(\phi/f)$ sets the global structure while $\cos(n\phi/f+\delta)$ sets the substructure of the potential.
One can easily check that when $cn^2\ll1$, the extrema of the potential are mainly set by $\cos(\phi/f)$ and thus appear only around $\phi/f=0,\pm\pi,\pm2\pi,\cdots$. In such a case the potential is effectively a
single cosine.
If, on the other hand, $cn\gg 1$, the positions of the extrema are determined by $\cos(n\phi/f+\delta)$.
Once the scalar starts oscillating along this potential, it will quickly get
trapped in one of the local minima of the substructure.

The case of interest for our purpose is 
\begin{equation}
 cn \ll 1 \ll cn^2,
  \label{cn-cond}
\end{equation}
which implies $ c \ll 1$ and $n \gg 1$.
In such a case the $\cos(n\phi/f+\delta)$ term in the
potential produces extrema,
but only within the distance of 
$\Delta \phi/f \sim cn$ from the extrema of $\cos(\phi/f)$.
Let us focus on this case and consider the dynamics of a homogeneous
scalar field along the two-cosine potential.

Assuming that the initial position of the scalar field~$\phi_*$ 
is located in the region $c n < \abs{\phi_*} / f \lesssim 1$,
then its dynamics at the beginning is set by the
$\cos(\phi / f)$ term in the potential, giving the scalar an
effective mass of~$m$.
Hence the field is initially frozen at~$\phi_*$ due to the Hubble
friction while $H > m$, and then starts to oscillate when $H \sim
m$~\cite{foot1}.

After the onset of the oscillation, it is useful to 
split the scalar density as
$\rho_\phi = \rho_{\mathrm{vac}} + \rho_{\mathrm{osc}}$,
where the first term denotes the vacuum energy $\rho_{\mathrm{vac}} =
V(\phi_{\mathrm{gl}})$ at the
global minimum~$\phi_{\mathrm{gl}}$ of the potential,
and the rest we refer to as the oscillation energy~$\rho_{\mathrm{osc}}$. 
The potential is well approximated by a quadratic except for the tiny
region within $|\phi| / f \lesssim cn$, hence
the oscillation is approximately harmonic and damped by the expansion of the
universe, so that $\rho_{\mathrm{osc}} \propto 1/a^3$.

\begin{figure}[b]
\centering  
  \includegraphics[width=0.85\linewidth]{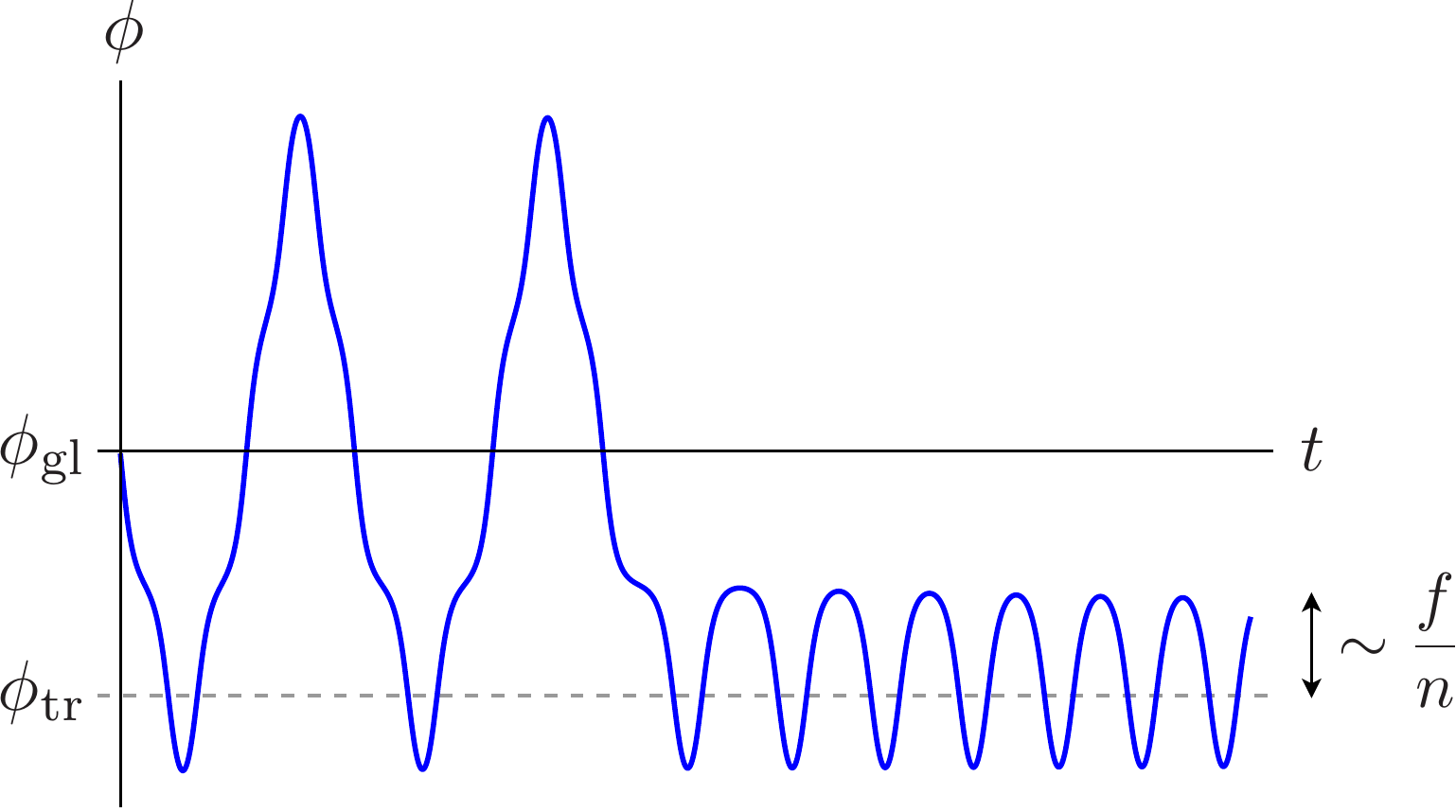}
  \caption{An example trajectory of a scalar field being trapped in a
 local minimum of the two-cosine potential~(\ref{2cos}).}
  \label{fig:trajectory}
\end{figure}

When the oscillation amplitude~$\bar{\phi}$ becomes sufficiently small,
$\bar{\phi} / f \sim cn$, the  
finer structure of the potential becomes relevant and the scalar field
eventually becomes trapped in one of the local minima created by the
$\cos (n \phi / f + \delta)$.
(It is also possible that the field gets trapped in the potential well
around the global minimum, however we do not
consider such a case.) 
The trapping minimum, which we represent by~$\phi_{\mathrm{tr}}$,
lies within the range of
\begin{equation}
 \frac{2 \pi }{n} \lesssim \frac{\abs{\phi_{\mathrm{tr}}}}{f} \lesssim c
  n.
  \label{phi_tr}
\end{equation}
Here the lower bound comes from the fact that 
each minimum is separated from their adjacent ones by 
$\Delta \phi / f \sim 2 \pi / n$,
and the assumption of $\phi_{\mathrm{tr}} \neq \phi_{\mathrm{gl}}$. 
It is not an easy task to obtain a general prediction of the value
of~$\phi_{\mathrm{tr}}$ within the above range. 
One may expect the field to be trapped at a minimum near
the upper end $\abs{\phi_{\mathrm{tr}}} / f \sim cn$,
however there the potential well is shallow and thus the trapping
probability is not necessarily high. 
Furthermore, $\phi_{\mathrm{tr}}$ is determined not just by the initial
condition~$\phi_*$, but also by the Hubble rate during trapping. 
In the following discussions we treat $\phi_{\mathrm{tr}}$ as a free
parameter within the range~(\ref{phi_tr}). 
An example of the scalar field trajectory upon trapping
for a case with $\abs{\phi_{\mathrm{tr}}} / f \sim 2 \pi / n$ 
is shown in Figure~\ref{fig:trajectory},
which we have obtained by numerically
computing the equations of motion.

After the trapping, the scalar's effective mass has increased
to $m_0 = \sqrt{V''(\phi_{\mathrm{tr}})} \sim c^{1/2} n \, m$,
while the field bound has decreased to $  f_0 = f/n$.
Hence the oscillation energy right after the trapping can be estimated as 
\begin{equation}
 \rho_{\mathrm{osc}}^+ \equiv
  \left. \rho_{\mathrm{osc}} \right|_{t = t_{\mathrm{tr}} + \epsilon}
  \sim m_0^2 f_0^2 \sim c m^2 f^2.
\label{rhooscp}
\end{equation}
Moreover, the trapping increases the vacuum energy by 
\begin{equation}
 \Delta \rho_{\mathrm{vac}} =
  V(\phi_{\mathrm{tr}}) -  V(\phi_{\mathrm{gl}})
  \sim \frac{1}{2} m^2  \phi_{\mathrm{tr}}^2 .
\label{Drhovac}
\end{equation}
By energy conservation the oscillation energy right before the trapping
is $ \rho_{\mathrm{osc}}^- \equiv
  \left. \rho_{\mathrm{osc}} \right|_{t = t_{\mathrm{tr}} - \epsilon}
  = \rho_{\mathrm{osc}}^+  + \Delta \rho_{\mathrm{vac}}  $,
hence the branching ratio of the oscillation energy into the vacuum
energy is
$\Delta \rho_{\mathrm{vac}}/\rho_{\mathrm{osc}}^- \sim 
\phi_{\mathrm{tr}}^2 / (\phi_{\mathrm{tr}}^2 + 2 c f^2).$
One sees that if $\abs{\phi_{\mathrm{tr}}} / f \sim 2 \pi / n$, then only
a tiny fraction of the oscillation energy is converted into vacuum energy.
On the other hand if $\abs{\phi_{\mathrm{tr}}} / f \sim c n$, most
of the oscillation energy goes into the vacuum~\cite{foot2}.

Viewing the oscillation energy~$\rho_{\mathrm{osc}}$ as the dark matter of our
universe, and the vacuum energy~$\rho_{\mathrm{vac}}$ as dark energy, the above
analyses suggest that the trapping happens when the dark matter density redshifts
down to $\rho_{\mathrm{DM}} = \rho_{\mathrm{osc}}^-$. Upon this `phase
transition', dark matter splits into a mixture of dark energy and
heavier dark matter, thus leading
to an increase in dark energy and a decrease in dark matter energy density. 

\noindent
\textit{Effective description of dark sector:}
For a scalar field that undergoes a trapping, the time evolution of its
energy density can be approximately described as,
\begin{equation}
 \rho_\phi (a)  = V(\phi_{\mathrm{gl}})
 + \Delta \rho_{\mathrm{vac}}
  \left(\frac{a + a_{\mathrm{tr}}}{a}\right)^{3}
  + \rho_{\mathrm{osc}}^+
  \left( \frac{a_{\mathrm{tr}}}{a} \right)^3,
\end{equation}
where $a_{\mathrm{tr}}$ is the scale factor at trapping,
$V(\phi_{\mathrm{gl}})$ is the vacuum energy (dark energy density) before the
trapping,
$\Delta \rho_{\mathrm{vac}}$ is the increase in the vacuum energy upon
trapping,
and $\rho_{\mathrm{osc}}^+$ is the oscillation energy (dark matter density) right
after trapping. 

For the two-cosine model~(\ref{2cos}), 
we have derived $\Delta \rho_{\mathrm{vac}}$ and $\rho_{\mathrm{osc}}^+$ 
in (\ref{Drhovac}) and (\ref{rhooscp}).
Supposing the trapping to have happened before today,
and the dark sector to consist entirely of the scalar field, 
then the present-day dark energy and dark matter densities are 
\begin{equation}
 \rho_{\mathrm{DE}0} \sim
  V(\phi_{\mathrm{g}}) +
 \frac{m^2 f^2}{2}  \left(\frac{\phi_{\mathrm{tr}}}{f}\right)^2,
\quad
\rho_{\mathrm{DM}0} \sim c\, m^2 f^2
\left(\frac{a_{\mathrm{tr}}}{a_0}\right)^3,
\label{DEDMtoday}
\end{equation}
where $a_0$ denotes the scale factor today.
Let us further estimate the trapping redshift by assuming
the initial field value as $\abs{\phi_*} \sim f$, and the oscillation
amplitude right before the trapping as
$\bar{\phi} \sim \abs{\phi_{\mathrm{tr}}}$;
then considering the oscillation prior to trapping to be mostly harmonic gives
$(\phi_{\mathrm{tr}} / f)^2 \sim (a_m / a_{\mathrm{tr}})^{3}$, with
$a_m$ being the scale factor at the onset of the oscillation
when $H \sim m$.
Here, $a_m$ can be computed by assuming the universe then to be
radiation-dominated, and also the 
entropy of the universe to be conserved thereafter.
Hence from the present-day entropy density, one can obtain the trapping
redshift as 
$ a_0 / a_{\mathrm{tr}} \sim
  10^{17} ( m / \mathrm{eV}  )^{1/2}
  ( \abs{\phi_{\mathrm{tr}}} / f )^{2/3}$.
(This result also depends on the number of relativistic degrees of freedom
at~$a_m$, however this dependence is weak and so it can be ignored.)

\noindent
\textit{Parameter space:}
Let us now assume that the dark energy before trapping is zero,
i.e. $V(\phi_{\mathrm{g}}) = 0$,
and see whether the present-day dark sector
can be explained by the two-cosine model.
There are effectively five free parameters $(c, n, m, f, \phi_{\mathrm{tr}})$,
out of which two are fixed by normalizing 
the dark sector densities~(\ref{DEDMtoday}) 
to their observed
values: 
$\rho_{\mathrm{DE} 0} \approx 3 \times 10^{-11} \, \mathrm{eV}^4$,
$\rho_{\mathrm{DM} 0} \approx 1 \times 10^{-11} \, \mathrm{eV}^4$~\cite{Ade:2015xua}.
We also remind the reader of the consistency conditions regarding the
trapping: (\ref{cn-cond}), (\ref{phi_tr}), and $a_{\mathrm{tr}} < a_0$.
In addition, the initial mass should be at least as large as
$m > 10^{-28}\, \mathrm{eV}$,
otherwise the scalar oscillation would not start by the matter-radiation
equality of the standard Big Bang cosmology.
Note also that the initial field value should be sub-Planckian,
$\abs{\phi_*} \sim f < M_{\rm Pl}$,
otherwise the scalar would drive (a secondary) inflation and dominate
the universe before staring its oscillation.

\begin{figure}[tb]
\centering  
  \includegraphics[width=0.85\linewidth]{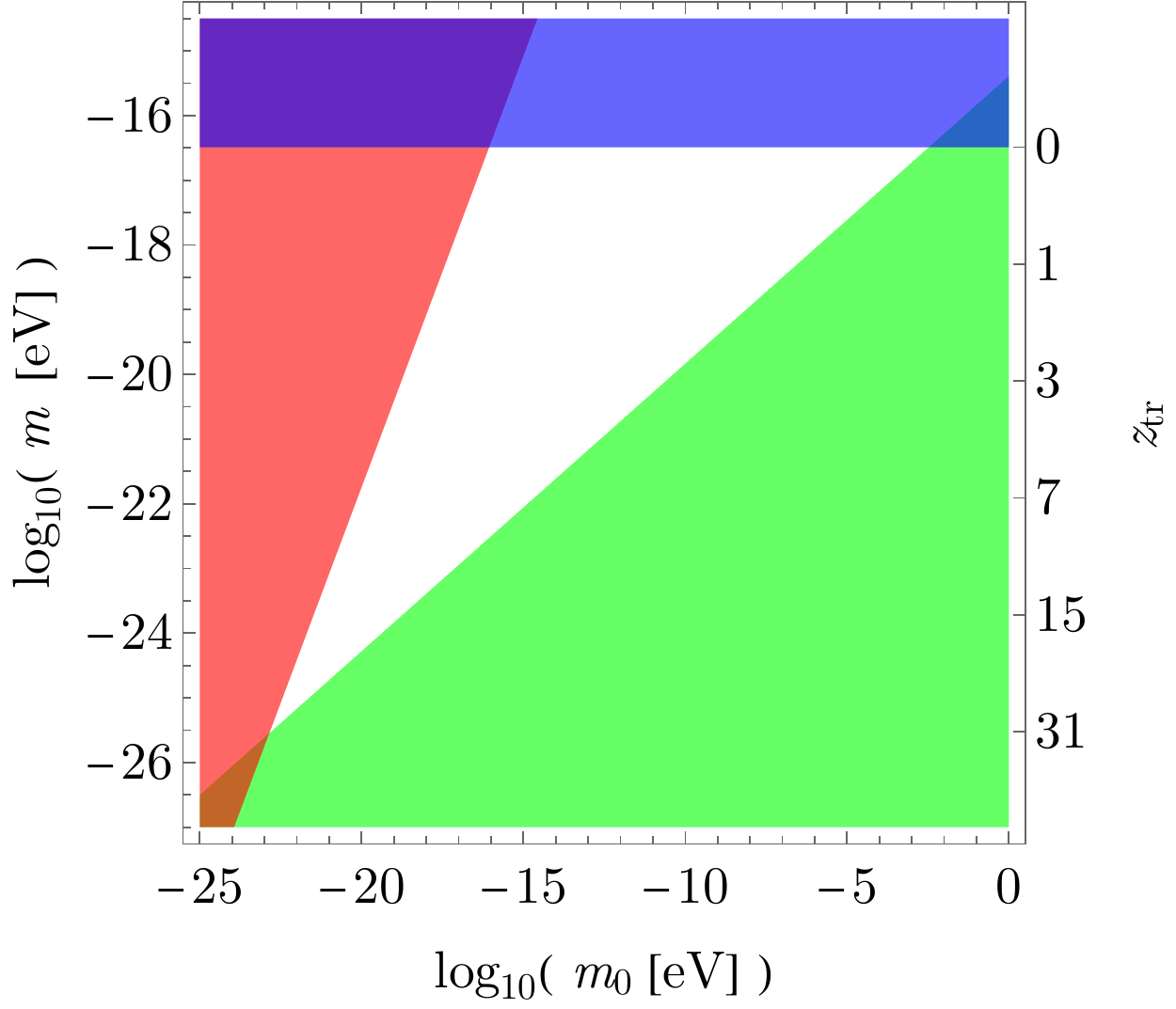}
 \caption{Dark matter mass before trapping ($m$) and after ($m_0$) that can
 explain the dark matter and dark energy of our universe, for the case of
 $f = 10^{16}\, \mathrm{GeV}$.
 The allowed region is shown in white, while the
 colored regions are excluded from the consistency conditions of
 $c n < 1$ (green), $\abs{\phi_\mathrm{tr}} / f > 2 \pi /  n$ (red),
 and $z_{\mathrm{tr}} > 0$ (blue).
 The right edge of the plot shows the redshift of the trapping.
 For a smaller~$f$, the allowed region shift towards larger values of
 the masses and redshift.}
  \label{fig:parspace}
\end{figure}

In Figure~\ref{fig:parspace} we show the allowed parameter window for a fixed
$f = 10^{16}\, \mathrm{GeV}$,
where the two-dimensional parameter space is displayed in terms of $m$ and $m_0$. 
Here the most restrictive constraints are the first inequality
of~(\ref{cn-cond}), the lower bound of~(\ref{phi_tr}), and
$a_\mathrm{tr} < a_0$.
These conditions exclude the parameter regions shown in green, red, and
blue, respectively.
The remaining window capable of explaining the observed dark matter and
dark energy densities is shown as the white region.
For fixed values of the dark sector densities and $f$,
the trapping redshift can be written as a function only of~$m$;
hence we also display $z_\mathrm{tr} = (a_0 / a_{\mathrm{tr}}) - 1$ on
the right edge of the plot.
If we further restrict ourselves to cases with trapping at 
$\abs{\phi_{\mathrm{tr}}} / f \sim 2 \pi / n$,
then we are on the boundary between the red and white regions.
Here the dark sector of our current universe can arise, for instance,
if the dark matter mass increases from $m \sim 10^{-21} \, \mathrm{eV}$
to $m_0 \sim 10^{-20} \, \mathrm{eV}$
upon trapping at $z_{\mathrm{tr}} \sim 5$;
the other parameters in this case are fixed to 
$c \sim 10^{-17}$,
$n \sim 10^{10}$, 
$f \sim 10^{16}\, \mathrm{GeV}$, and 
$f_0 \sim 10^6\, \mathrm{GeV}$.
Trapping always happens after the matter-radiation equality for $f =
10^{16}\, \mathrm{GeV}$; however with decreasing~$f$,
the allowed regions for the dark matter masses as well as the trapping
redshift~$z_{\mathrm{tr}}$ tend to shift towards larger values.
In particular if $ f \lesssim 10^{10}\, \mathrm{GeV}$, the trapping can happen at
times before the equality.
We should also remark that recent studies of the Lyman-$\alpha$ forest have
constrained the mass of scalar dark matter to be larger than about $10^{-21}\,
\mathrm{eV}$~\cite{Irsic:2017yje,Armengaud:2017nkf,Kobayashi:2017jcf}. 
However this bound does not directly apply to our~$m$,
since the Lyman-$\alpha$ analyses assume the dark matter mass to be
time-independent after the equality.

We have also checked the stability of the trapped minimum against quantum tunneling,
focusing on cases where the scalar is trapped in a local minimum
(false vacuum) adjacent to the global minimum (true vacuum)
and computing the Coleman-De~Luccia tunneling rate~\cite{Coleman:1980aw}.
For the parameters in Figure~\ref{fig:parspace}, the lifetime of the
trapped vacuum is much longer than the age of the universe,
which is basically due to the energy density difference between the false
and true vacua being normalized to the dark energy, and therefore tiny.

\noindent
\textit{`Axion' abundance:}
After the trapping, the oscillating scalar can be interpreted as a
collection of axion-like particles with mass~$m_0$ and ``axion decay
constant''~$f_0$ 
(although this is an abuse of language as 
in the toy model under study we do not consider any
direct couplings between $\phi$ and other matter fields.)
Using these quantities, the ratio of the oscillation energy to the
critical density today can be written as 
\begin{multline}
 \Omega_{\mathrm{osc}} h^2 \sim 10^{-1} 
  \left(\frac{m_0}{10^{-22}\, \mathrm{eV}}\right)^{1/2}
 \left(\frac{f_0}{10^{17}\, \mathrm{GeV}} \right)^2
 \\
 \times \frac{(c n^2)^{11/4}}{(cn)^2}
 \left(\frac{\phi_{\mathrm{tr}}}{f_0}\right)^{-2}.
\label{Omega_osc}
\end{multline}
If the second line is ignored, this expression is exactly the same as
for the traditional axion-like particles with initial field
displacement $\abs{\phi_*} \sim f_0$
(see, e.g., Eq.~(3.10) of~\cite{Kobayashi:2017jcf}).
However the trapping gives rise to the second line, which is guaranteed
to be larger than unity from (\ref{cn-cond}) and (\ref{phi_tr}).
This enhancement is understood by noting that
the traditional axion density starts to redshift from its initial
value~$m_0^2 f_0^2$ when the axion begins to oscillate at $H \sim m_0$.
On the other hand, the two-cosine scalar begins to oscillate at a later time
when $H \sim m (< m_0)$, and then after a while it gets trapped;
it is at this trapping time that the oscillation energy becomes of~$m_0^2 f_0^2$, 
cf.~(\ref{rhooscp}).
(See \cite{DAmico:2016jbm,Jaeckel:2016qjp} for related discussions in
the context of monodromy dark matter.)

\noindent
\textit{Inhomogeneities:}
Thus far we have focused on the scalar dynamics of the
homogeneous background. However it is also important to consider the
inhomogeneities, particularly because inflation produces scalar
field fluctuations on super-horizon scales given that the scalar existed
during inflation and the inflationary
Hubble rate~$H_{\mathrm{inf}}$ was greater than~$m$.
The field fluctuations give rise to isocurvature perturbations in the dark
sector~\cite{Linde:1985yf,Seckel:1985tj,Lyth:1989pb,Kobayashi:2013nva}.
However we note that the existing limits on isocurvature are mainly from 
measurements of the cosmic microwave background (CMB),
which constrains dark matter isocurvature at recombination.
Hence if the trapping happens at a later time, the dark matter isocurvature
at recombination was
$\delta \rho_{\mathrm{DM}} / \rho_{\mathrm{DM}} \sim H_{\mathrm{inf}} /
(2 \pi f)$,
and thus the isocurvature measured by CMB would be much
smaller than what one would naively guess from the present-day decay constant~$f_0$.
This feature of the trapping, together with the enhancement of the axion
abundance~(\ref{Omega_osc}),
allows axion-like particles to evade the various standard cosmological
consistency relations.

We should also remark that upon trapping, the inhomogeneities may grow
as the scalar oscillates along the potential with
substructure~\cite{Jaeckel:2016qjp}, which may even lead to formation of 
oscillons~\cite{Amin:2011hj}. 
Moreover, the initial field fluctuation may induce the scalar to be
trapped in different local minima in different patches of the universe. 
This would lead to formation of domain walls, which are likely to
annihilate each other due to the energy density difference between the
various vacua. 
These walls may not disappear by today, but they do not necessarily 
dominate the universe if the trapping happened at a low redshift.
Moreover, inhomogeneous trapping gives rise to inhomogeneous dark energy.
We also mention that in regions of the universe where the dark
matter density is high, such as inside galaxies, the scalar may be
untrapped and oscillate along the global potential,
leading to dark matter properties different from those in the
intergalactic space.
All these features can provide smoking-gun signals of the scenario.

\noindent
\textit{Conclusions:}
Let us then recap and summarize the broad features of this model. If we first focus on the dark matter associated with the
emergent dark energy, we can see that it was lighter in the past, with
density larger compared to a naive 
extrapolation from its present-day value.
In particular if the trapping happened between recombination and today,
this would give rise to apparent discrepancies between cosmological
measurements using the CMB and low-redshift probes.
From this point of view, it would be very interesting to study the implications of our
scenario for the recent tensions in the measurements of $\sigma_8$ and
$H_0$~\cite{Battye:2014qga,Freedman:2017yms}. 
We also note that if the initial dark matter mass was ultralight,
this will have an effect on
structure formation with a greater suppression of small scale structure
in the past as compared to today.
Dark matter becoming heavier can also shorten its lifetime, and thus may
lead to enhanced signals in indirect searches.
Furthermore, when viewing the dark matter
as a collection of axion-like particles, 
we showed that naive
estimates of the abundance and isocurvature modes
based on the axion's current mass and decay constant 
will most likely be wrong---we expect a larger abundance
today, as well as a far smaller isocurvature mode arising at early times.  

With regards to the dark energy, we find that it was smaller
in the past. There are two possibilities that should be
considered. The first is that the true, global, minimum (or minima) of
the potential are exactly zero. Then, the fact that trapping minima generically
occur near the global minima would
naturally lead to a small cosmological constant, in the sense that the field would be
accidentally caught in a wrinkle close to where $V\simeq0$.
In this picture, the question of why the dark matter and dark energy
densities are of the same order today, can be rephrased as, why did the
trapping happen at the right time?
From this point of view, it would be important to 
explore microscopic realizations of the trapping to see how the 
trapping time is constrained in explicit models.
(A multi-cosine model  may be
constructed using the clockwork mechanism~\cite{Choi:2015fiu,Kaplan:2015fuy}.)
Another possibility is that the global minima are negative and so
the global vacuum structure is anti-DeSitter space. 
In this
case, the complexity of the potential would be one way of explaining why
we could live in such a universe with what is effectively a positive
cosmological constant. 
We also note that the increasing dark energy, when averaged over time,
gives rise to an equation of state~$w < -1$; such a behaviour is
preferred by some recent observations~\cite{Shafer:2013pxa}.

In this letter we have presented an intriguing possibility, that complex
potential might lead to a small cosmological constant
from an energy difference between its global and local minima, 
and that dark energy and dark matter might be
intertwined. A more thorough analysis is required to check if this is 
truly viable, i.e. if it leads to the distances and growth rates which
are consistent with current observations. We leave this for future work.

\begin{acknowledgments}
We thank Aleksandr Chatrchyan, Joerg Jaeckel, Kazunori Kohri, Viraf
Mehta, Lorenzo Ubaldi, and Matteo Viel for helpful discussions,
as well as the anonymous referee for useful comments.
TK also thanks the Department of Physics at University of Oxford for
hospitality while this work was in progress,
and acknowledges support from the Sciama Legacy Bursary and INFN INDARK PD51 grant.
PGF acknowledges support from  STFC, the Beecroft Trust and the ERC. 
\end{acknowledgments}

\appendix



\begin{thebibliography}{99}
\bibitem{foot1} 
The second derivative of the potential is dominated by 
the $\cos (n \phi / f + \delta)$ term, i.e., $\abs{V''} \sim cn^2 m^2$.
However, even during the `frozen' phase, the
scalar is actually slowly rolling. Therefore the time-averaged second
derivative from around the onset of the oscillation
is $\langle V'' \rangle \sim m^2 $.
\bibitem{foot2} 
We should
comment on the special cases where the calculations break down.
If the scalar is trapped  
very close to the edge of the region with minima,
$\abs{\phi_{\mathrm{tr}}} / f \approx cn $,
then the potential well is very shallow and hence
$\rho_{\mathrm{osc}}^+$ becomes much smaller than the estimate
of~(\ref{rhooscp}). 
On the other hand if $\delta \approx \pi$, and further if the scalar is trapped
in a local minimum that is almost degenerate with the global one, then
$\Delta \rho_{\mathrm{vac}}$ becomes much smaller than the estimate
of~(\ref{Drhovac}).
\end{thebibliography}

\end{document}